
\def\chaphead{}
\def\ni{\noindent}

\font\tfont=cmbxti10
\font\eightrm=cmr8
\font\eightit=cmti8
\font\sixrm=cmr6
\font\eightmit=cmmi8
\font\sixmit=cmmi6
\def\absmath{\textfont0=\eightrm \scriptfont0=\sixrm
	      \textfont1=\eightmit \scriptfont1=\sixmit}
\def\absfont{\let\rm=\eightrm \let\it=\eightit \rm\absmath}
\font\twelverm=cmr12
\font\twelveit=cmti12
\font\tenrm=cmr10
\font\twelvemit=cmmi12
\font\tenmit=cmmi10
\def\regmath{\textfont0=\twelverm \scriptfont0=\tenrm
	      \textfont1=\twelvemit \scriptfont1=\tenmit}
\def\peterfont{\let\rm=\twelverm \let\it=\twelveit \rm\regmath}
%
%

\newfam\vecfam

\textfont\vecfam=\tfont \scriptfont\vecfam=\seveni
\scriptscriptfont\vecfam=\fivei


\def\spose#1{\hbox to 0pt{#1\hss}}

\font\eightrm=cmr8

\def\s{\ifmmode \widetilde \else \~\fi} 
     
\def\section{\S}
\newcount\notenumber
\notenumber=1
\newcount\eqnumber
\eqnumber=1
\newcount\fignumber
\fignumber=1
\newbox\abstr


\def\s{{\rm\,s}}

\def\note#1{\footnote{$^{\the\notenumber}$}{#1}\global\advance\notenumber by 1}
\def\foot#1{\raise3pt\hbox{\eightrm \the\notenumber}
     \hfil\par\vskip3pt\hrule\vskip6pt
     \noindent\raise3pt\hbox{\eightrm \the\notenumber}
     #1\par\vskip6pt\hrule\vskip3pt\noindent\global\advance\notenumber by 1}

\def\abstract#1{\setbox\abstr=\vbox{\hsize 5.0truein{\par\noindent#1}}
    \centerline{ABSTRACT} \vskip12pt \hbox to \hsize{\hfill\box\abstr\hfill}}
     
\def\Dt{\spose{\raise 1.5ex\hbox{\hskip3pt$\mathchar"201$}}}    
\def\dt{\spose{\raise 1.0ex\hbox{\hskip2pt$\mathchar"201$}}}    

\def\new{{\rm\chaphead\the\eqnumber}\global\advance\eqnumber by 1}
\def\ref#1{\advance\eqnumber by -#1 \chaphead\the\eqnumber
     \advance\eqnumber by #1 }
\def\last{\advance\eqnumber by -1 {\rm\chaphead\the\eqnumber}\advance
     \eqnumber by 1}
\def\eqnam#1{\xdef#1{\chaphead\the\eqnumber}}
     
\def\nfig{\chaphead\the\fignumber\global\advance\fignumber by 1}
\def\nfiga#1{\chaphead\the\fignumber{#1}\global\advance\fignumber by 1}
\def\rfig#1{\advance\fignumber by -#1 \chaphead\the\fignumber
     \advance\fignumber by #1}
\def\fignam#1{\xdef#1{\chaphead\the\fignumber}}

\def\lta{\mathrel{\spose{\lower 3pt\hbox{$\mathchar"218$}}
     \raise 2.0pt\hbox{$\mathchar"13C$}}}
\def\gta{\mathrel{\spose{\lower 3pt\hbox{$\mathchar"218$}}
     \raise 2.0pt\hbox{$\mathchar"13E$}}}
     

\magnification=\magstep1
\parskip=3pt

\magnification=\magstep1
\baselineskip=14pt
\def\ni{\noindent}
\def\op{\Omega_p}
\def\os{\Omega_*}
\def\refind{\noindent \hangindent=2pc \hangafter=1}

\centerline{ }
\centerline{\bf Differential rotation enhanced dissipation of tides in
 the PSR J0045-7319 Binary}
\bigskip
\centerline{Pawan Kumar$^\dagger$ and Eliot J. Quataert}
\medskip
\centerline{Department of Physics}
\medskip
\centerline{Massachusetts Institute of Technology, Cambridge, MA 02139}
\bigskip

\centerline{\bf Abstract}
\bigskip
Recent observations of PSR J0045-7319, a radio pulsar in a close
eccentric orbit with a massive B-star companion, indicate that the
system's orbital period is decreasing on a timescale of $\approx 5
\times10^{5}$ years, which is much shorter than the timescale of
$\approx$ $10^9$ years given by the standard theory of tidal
dissipation in radiative stars. Observations also provide strong
evidence that the B-star is rotating rapidly, perhaps at nearly its
break up speed. We show that the dissipation of the dynamical tide in
a star rotating in the same direction as the orbital motion of its
companion (prograde rotation) with a speed greater than the orbital
angular speed of the star at periastron results in an increase in the
orbital period of the binary system with time.  Thus, since the
observed time derivative of the orbital period is large and negative,
the B-star in the PSR J0045-7319 binary must have retrograde rotation
if tidal effects are to account for the orbital decay.  We also show
that the time scale for the synchronization of the B-star's spin with
the orbital angular speed of the star at periastron is comparable to
the orbital evolution time. From the work of Goldreich and Nicholson
(1989) we therefore expect that the B-star should be rotating
differentially, with the outer layers rotating more slowly than the
interior. We show that the dissipation of the dynamical tide in such a
differentially rotating B-star is enhanced by almost three orders of
magnitude leading to an orbital evolution time for the PSR J0045-7319
Binary that is consistent with the observations.

\vskip 3.0truecm
\noindent $^\dagger$ Alfred P. Sloan Fellow \& NSF Young Investigator

\noindent email addresses: pk@sns.ias.edu (P. Kumar)

\hskip 55pt equataert@cfa.harvard.edu (E.J. Quataert)

\vfill\eject
\centerline{\bf 1. Introduction}
\bigskip

\baselineskip=15pt
An unusual binary system was recently discovered in the Small
Magellanic Cloud (SMC); Johnston et al. 1994, Kaspi et al. 1994. One
of the members of this binary is a pulsar, PSR J0045-7319, with a spin
period of 0.93 s; its companion is a main sequence B-star of mass
$\sim 8.8 M_\odot$.  The orbital period of this binary system is 51
days, the eccentricity is 0.808, and the separation between the stars
at periastron is about 4 times the B-star radius.  Kaspi et al. (1996)
report that the orbital period for this system is evolving on a time
scale of $\sim 5 \times 10^5$ years, whereas the standard theory for
tidal dissipation predicts an evolution time of order $10^9$
years. Observations also indicate that the spin of the B-star is not
perpendicular to the orbital plane and that the component of the spin
normal to the orbital plane is close to the break up rotation speed
(Kaspi et al. 1996, Lai et al. 1995).

It has been suggested (Lai 1996) that the rapid evolution of the orbital
period of the SMC binary can be accounted for by the radiative
dissipation of the dynamical tide, provided that the B-star companion
to PSR J0045-7319 has a rapid retrograde rotation rate. This is an
interesting idea and the observations do suggest that the B-star is
rapidly rotating, though the sign of the spin is unknown.  We show,
however, that as long as the B-star is rotating rigidly, the
dissipation of the dynamical tide is too inefficient to explain the
observed evolution of the orbital period.  When the star has some
differential rotation, so that the frequency of the tidally excited
waves in the rest frame of the fluid near the surface of the star is
smaller than in the deep interior, the wave dissipation increases
dramatically, readily explaining the observations.  A more detailed
and rigorous presentation of this analysis will be given in a future
paper (Kumar \& Quataert 1996).

\bigskip
\centerline{\bf 2. The dynamical tide in a rigidly rotating star }
\bigskip

Waves excited by the tidal forcing have frequencies $\approx 2 \op$ as
seen by an inertial observer, where $\op$ is the angular speed of the
star at periastron. If the angular velocity of the B-star
perpendicular to the orbital plane is
$\os$, then the wave frequency in the local rest frame of the fluid
is $\omega_{t*}\approx 2[\op-\os]$.  The angular momentum
luminosity in the dynamical tide (the angular momentum transported by
waves across a sphere of radius $r$ per unit of time) is

$$ \dot L_{tide} = 2 \rho r^2  \omega_{t*} \xi_t^2(r) v_g,
\eqno(\new) $$ where $\xi_t^2$ is the square of the transverse
displacement (integrated over solid angle) associated with the
dynamical tide, $v_g\sim r \omega_{t*}^2/(6^{1/2} N)$ is the group
speed of quadrupole gravity waves in the limit of small wave
frequencies ($\omega_{t*}\ll N$), and $N$ is the
Brunt-V\"ais\"al\"a~frequency. We note that the radial displacement
amplitude of the dynamical tide is much smaller than its transverse
displacement and therefore has been neglected in the above equation.
It can be shown (Zahn 1975) that $\xi_{t}$ is approximately equal to the
transverse displacement associated with the equilibrium tide,
$\xi_t^{(eq)}$, about a wavelength away from the convective core.
Furthermore, it is easy to show that the squared transverse
displacement of the equilibrium tide integrated over all angles is

$$ \left|\xi_t^{(eq)}\right|^2 = {3\pi\over 10} {G^2 M^2\over R^6}
\left[ {d\over dr} \left({r^4\over g}\right)\right]^2, \eqno(\new)$$
where $R$ is the distance between the two stars, $M \approx 1.4
M_\odot$ is the mass of the neutron star, and $g$ is the gravitational
acceleration of the primary star. The total angular momentum in the 
dynamical tide is
given by $L_{tide} \approx T_{peri} \dot L_{tide}$, where $T_{peri}
\approx \pi(1-e)^2/(\Omega_o\sqrt{1-e^2}) \approx 2$ days is the duration 
of the periastron encounter and $e = 0.808$ is the orbital
eccentricity.  The rate of change of the orbital energy due to the
dissipation of the dynamical tide is $\dot E_{orb} = -\Gamma
E_{tide}$, where $\Gamma$ is the dissipation rate for the dynamical
tide (calculated below) and $E_{tide}$ is the energy in the dynamical
tide, which can be shown to be equal to $L_{tide}\op$, i.e., $$
E_{tide} \approx {6\pi^2\over 600^{1/2}}
\left({M\over M_t}\right)^2 \left[{\omega_{t*}^3 \Omega_o^4\over
(1-e)^6 }\right] {r\rho\over N} \left[ {d\over dr}\left({r^4\over g}
\right)\right]^2,
\eqno(\new)$$
where $M_t \approx 10.2 M_\odot$ is the total mass of the binary
system, $\Omega_o=(GM_t/a^3)^{1/2} = 1.42 \times 10^{-6}$ rad s$^{-1}$
is the orbital frequency and $a\approx 126 R_\odot$ is the semi-major
axis of the orbit.  Note that all quantities in the above equation
that depend on $r$ are evaluated about one wavelength away from the
inner turning point of the wave.  We take the radius of the B-star to
be $6.0 R_\odot$; the radius of the convective core of the
B-star, where the gravity waves become evanescent, is about 0.23
$R_*$, the mass in the convective core is $\approx 3 M_\odot$ and the
density and Brunt-V\"ais\"al\"a~frequency outside the core are 2 g
cm$^{-3}$ and 100 $\mu$Hz, respectively.  Substituting these into
equation (\last) we find that the total energy in the dynamical tide
for a non-rotating star, so that $\omega_{t*} \approx 2\op \approx 7.2
\mu$Hz, is $\approx 10^{40}$ ergs.  For retrograde rotation of the
B-star at a frequency of $\Omega_*/2\pi= -8 \mu$Hz, or $\hat \Omega_*
\equiv
\Omega_*(R_*^3/GM_*)^{1/2}
\approx -0.4$, the frequency of
the tidally excited wave is $\omega_{t*}\approx 23 \mu$Hz and equation
(\last) implies that the energy in the tide is a factor of about 100
larger compared to the non-rotating star. This point was made by
Lai (1996) who suggested that the enhanced energy in the dynamical tide
can explain the observed evolution of the SMC system. Unfortunately,
as we point out below, this is not tenable because the damping time of
a wave of frequency $\approx 23\mu$Hz is also larger by a factor of
about 100 than a wave of frequency $\approx 7.2\mu$Hz. 

For prograde rotation the tidal frequency and energy decrease with
increasing rotation rate so long as $\os<\op$ (see eq. [3]).  The
tidal energy does not actually vanish when $\os=\op$ because
axisymmetric waves are still excited.  For $\os>\op$ the tidal torque
causes the angular momentum of the stellar spin to decrease (which is
taken up by the orbit) and as a result the orbital energy and the
orbital period increase with time. Equation (3) shows that $E_{tide} <
0$, and so $\dot E_{orb} > 0$, when $\os>\op$, which may seem
paradoxical. We note that the tidal energy as seen in the rotating
frame of the star is a positive definite quantity; the tidal energy as
seen from an inertial frame (given in eq. [\last]), however, also
includes a contribution from the work done by the tidal force on the
rotating fluid in the star, which can either increase or decrease the
star's rotational kinetic energy. The negative tidal energy for $\os >
\op$ corresponds to more energy being taken out of the rotation than
is deposited in the dynamical tide. We emphasize that the tidal energy
is positive so long as $\os<\op$, which obviously includes the case of
arbitrary retrograde rotation, and is negative when $\os>\op$. Thus,
if the rotation rate of the B-star is greater than $\op$, the observed
period decrease of the PSR J0045-7319 binary implies that the direction
of the rotation of the B-star must be retrograde. We also note that
the energy in the tide for $\os\approx\op$ is too small by about 2
orders of magnitude to explain the observed rate of period
evolution.

A detailed calculation of the damping of gravity waves is given in
Kumar \& Quataert (1996). Here we present a crude estimate to point out
that the wave damping increases very rapidly with decreasing wave
frequency.

The local damping rate of a wave, $\gamma(r)$, is approximately equal
to the inverse of the time it takes photons to diffuse across a
wavelength times the ratio of the photon energy density to the total
thermal energy density of the plasma.  This can be written in the
following convenient form, $$\gamma(r)\sim F k_r^2/(\rho g),
\eqno(\new)$$ where $F$ is the radiative flux of the star and $k_r\sim N
\sqrt{\ell(\ell+1)}/(r\omega_{t*})$ is the radial wave number for gravity 
waves ($\ell = 2$ for quadrupole waves).  We see from the above
expression that the wave damping increases with decreasing density in
the star and thus most of the contribution to the wave damping comes
from near the upper turning point of the wave which is located at a
depth of $\sim R^2_* \omega^2_{t*}/(6 g)$ beneath the stellar surface.
Integrating the local dissipation rate, weighted by the wave's energy
density, yields the global wave damping rate. Since the energy density
in the wave is inversely proportional to its group speed, we find that
the global damping rate for the wave, in a star of polytropic index 2
(a value appropriate for the outer envelope of the Yale group's B-star
model we are using), scales as $\omega^{-7}_{t*}$. This scaling only
applies to waves with more than about 4 radial nodes, which
corresponds to $\omega_{t*} \lta 15 \mu$Hz for the SMC B-star. Modes
with fewer nodes or higher frequencies have wavelengths of order the
stellar radius and their dissipation is not as concentrated near the
outer turning point as it is for lower frequency modes. Thus the
global damping rate of high frequency g-modes has a much weaker
dependence on mode frequency. Using equation (4) and the parameters
for the SMC B-star given above we estimate that the damping time for a
low order quadrupole gravity wave is $\sim 3000$ years, which is
consistent with the results of a number of different nonadiabatic
eigenvalue calculations (Kumar \& Quataert 1996; Saio \& Cox 1980).

The frequency of the dynamical tide in the non-rotating star is
$\approx 7.2 \mu$Hz, while for rapid retrograde rotation
($\hat\Omega_* = -0.4$) the frequency is $\approx 23 \mu$Hz.  Using
the above scaling for the damping of the dynamical tide, we find that
the ratio of the damping times in these two cases is $\approx 1/200$.
Since the rate of dissipation of the orbital energy is equal to the
tidal energy times its damping rate, we see that even
though the energy in the tide for rapid retrograde rotation is a
factor of $\sim 100$ larger than for no rotation, the rate of
dissipation of the orbital energy is very nearly the same for
$\hat\Omega_* = -0.4$ and $\hat \Omega_* = 0$.  Using the dissipation
time for the dynamical tide in the retrograde rotating star of $\sim
3000$ years, the energy in the dynamical tide of $\sim 10^{42}$ ergs,
and the orbital energy for the SMC binary, $E_{orb} \approx
2\times10^{47}$ ergs, we find that the orbital period evolution time
for the SMC Binary is $\approx 10^9$ years, or roughly three orders of
magnitude larger than the observed value.

\bigskip
\centerline{\bf 3.Differential rotation enhanced dissipation of tides }
\bigskip

It should be clear from this discussion that one way to get more rapid
evolution of the orbit is to find a more efficient way of dissipating
the energy in the dynamical tide. This would occur naturally,
considering the sensitivity of wave dissipation on frequency, if the
frequency of the tidally excited waves were smaller near the surface
of the star, where the dissipation occurs.  Waves with frequencies
$\approx 2\op$, as seen in an inertial reference frame, are excited
near the interface of the convective core and the radiative exterior
where the wavelength of the wave is the largest.  Thus, in order to
get a smaller wave frequency in the rest frame of the fluid near the
surface of the star, we need the surface to be rotating at a frequency
close to $\Omega_p \approx 3.6 \mu$Hz.  We show below that this is in
fact expected for the B-star of the SMC binary.

For nearly circular orbits, the rotation of stars in close binary
systems tends to synchronize with the orbital motion, that is, the
rotation frequency of the star approaches the orbital frequency.  For highly
eccentric orbits, the tidal force is only appreciable near periastron
and so the rotation frequency of the star approaches $\Omega_p$.
This is called pseudo-synchronization.

We estimate the relative timescales for spin pseudo-synchronization
and orbital circularization for the SMC system to determine if the B-star
should be rotating pseudo-synchronously.  The pseudo-synchronization
timescale due to the transfer of orbital angular momentum to the star
is $L_*/\dot L_* = -(L_*/L_{orb})(L_{orb}/\dot L_{orb})$, where
$L_{orb}/L_* \approx 4 \hat \Omega^{-1}_*$ for the SMC Binary and
$\dot L_{orb} \approx \dot E_{orb}/\op$ (see above).  Since $L_{orb}
\Omega_o/E_{orb} \approx 1$, we find that $L_*/\dot L_* 
\approx  5 \hat \os E_{orb}/\dot E_{orb}$ for the SMC Binary system.  
It follows from the general equations for tidal evolution in the weak
friction limit that a similar result holds for the equilibrium
tide (Hut 1981). We conclude that so long as the B-star in the SMC Binary
is rotating near break up, both the equilibrium and the dynamical
tides lead to timescales for spin pseudo-synchronization which are
comparable to the orbital circularization time.  We therefore do not
expect that the B-star's interior has spun down appreciably, which is
clearly consistent with the observations.

We do expect, however, that the surface of the B-star will have slowed
down, leading to appreciable differential rotation in the B-star. The
physical reason for this follows from the seminal work of Goldreich
and Nicholson (1989). Waves generated by tidal forces deposit their
angular momentum at the place in the star where they are dissipated.
Since the dissipation rate is largest near the surface of the star,
and the moment of inertia of the surface region is a small fraction of
the star's total moment of inertia, the outer region of the star tends
to be pseudo-synchronized on a much shorter time scale than the
interior.  For the retrograde rotating B-star with $\hat \os \approx
-0.3$, the frequency of the dynamical tide is $\approx 17
\mu$Hz; the outer turning point of this wave is at $r \approx 0.9 R_*$ and
so the rotation of the B-star at $r \approx 0.9R_*$ should be
pseudo-synchronous.  Those parts of the star lying above or
significantly below this radius, however, experience little net tidal
torque and thus can continue to rotate at the primordial rotation rate
of the star provided that magnetic stresses and shear instabilities do
not efficiently transport angular momentum to/from the
pseudo-synchronous layer. The optical linewidth measurement of Bell et
al. (1995) gives a projected rotation velocity of the B-star of
113$\pm$10 km s$^{-1}$.  If correct, this suggests that the surface of
the star is perhaps not pseudo-synchronized, supporting our simple
model.

The frequency of the tidally excited gravity wave in the local rest
frame of the star a distance $r$ from the center is $\omega_{t*}(r)
\approx 2[\op - \Omega_*(r)]$ and its wavenumber is $\sim
6^{1/2}N(r)/(r\omega_{t*})$, where $\os(r)$ is the local angular 
rotational velocity of the star normal to the orbital plane.
Thus, as the wave approaches the
pseudo-synchronously rotating surface of the star its frequency and
wavelength approach zero and the energy it carries is entirely
dissipated. The energy in the dynamical tide for retrograde stellar
rotation at a frequency of $\sim -6 \mu$Hz ($\hat
\Omega_* \approx -0.3$) is $\sim$ 10$^{41}$ ergs (see eq. [3]). 
With a change in the orbital energy per orbit of $\Delta E_{orb}
\approx - E_{tide} \approx - 10^{41}$ ergs, and the orbital energy of
$2\times 10^{47}$ ergs, we find that $P_{orb}/\dot P_{orb} = 1.5
E_{orb}/\dot E_{orb}\approx 4.2\times10^5$ years, which in good
agreement with the observations. It is easy to show that the rate of change
of orbital eccentricity ($\dot e$) is a factor of about
7 smaller than $\dot P_{orb}/P_{orb}$. Thus the theoretically expected
value for $\dot e$ is a factor of $\sim 20$ smaller than the current
observational limit (Kaspi et al. 1996).

We note that prograde rotation of the
B-star at $\hat \Omega_* \approx 0.4$ gives an orbital evolution time
of comparable magnitude, though with the opposite sign.  This
emphasizes that retrograde rotation is not required to explain the
observed short timescale for orbital evolution, but rather to get the
correct sign for the time derivative of the orbital period.

\vfill\eject
\bigskip
\centerline{\bf REFERENCES}
\bigskip

\refind Bell, J.F., Bessell, M.S., Stappers, B.W., Bailes, M., Kaspi, V.M. 1995, ApJLetters, 447, 117

\ni Goldreich, P. and Nicholson, P. D. 1989, ApJ, 342, 1079

\ni Hut, P. 1981, A \& A, 99, 126

\refind Johnston, S., Manchester, R.N., Lyne, A.G., Nicastro, L., and
    Spyromilio, J., 1994, MNRAS, 268, 430

\refind Kaspi, V.M., Baile, M., Manchester, R.N., Stappers, B.W., and Bell, 
  J.F. 1996, Nature, 381, 584

\refind Kaspi, V.M., Johnston, S., Bell, J.F., Manchester, R.N., Bailes, M., Bessell, M., Lyne, A.G., and D'amico, N., 1994, ApJ, 423, L43

\ni Kumar, P. \& Quataert, E., 1996, ApJ (in preparation)

\ni Lai, D. 1996, ApJ, 466, L35

\ni Lai, D., Bildsten, L., and Kaspi, V.M 1995, ApJ, 452, 819

\ni Saio, H. and Cox, J.P. 1980, ApJ, 236, 549

\ni Zahn, J-P, 1975, Astr. Ap., 41, 329
\bye